\documentclass[%nofootinbib
 reprint,
superscriptaddress,
 amsmath,amssymb,
 aps,
 prl,
floatfix]{revtex4-2}

\usepackage{graphicx}% Include figure files
\graphicspath{{images/}}
\usepackage{dcolumn}% Align table columns on decimal point
\usepackage{bm}% bold math
\newcommand{\Br}[1]{\ensuremath{\left( #1 \right)}} %Matched parentheses.
\usepackage{amssymb}
\usepackage{soul}
\usepackage[dvipsnames]{xcolor}
\usepackage[normalem]{ulem}

\begin{document}

\preprint{APS/123-QED}
%%TC:ignore
\title{Mechanical squeezing via fast continuous measurement}% Force line breaks with \\

\author{Chao Meng}
\thanks{These authors equally contributed to this Letter.}
\author{George A. Brawley}%
\thanks{These authors equally contributed to this Letter.}
\author{James S. Bennett}
\affiliation{%
	Australian Research Council Centre of Excellence for Engineered Quantum Systems, School of Mathematics and Physics, University of Queensland, St Lucia, Queensland 4072, Australia.
}%
\author{Michael R. Vanner}
\affiliation{%
	Australian Research Council Centre of Excellence for Engineered Quantum Systems, School of Mathematics and Physics, University of Queensland, St Lucia, Queensland 4072, Australia.
}
\affiliation{QOLS, Blackett Laboratory, Imperial College London, London SW7 2BW, United Kingdom}
\author{Warwick P. Bowen}

\email{wbowen@physics.uq.edu.au}
\affiliation{%
 Australian Research Council Centre of Excellence for Engineered Quantum Systems, School of Mathematics and Physics, University of Queensland, St Lucia, Queensland 4072, Australia.
}%

\date{\today}% It is always \today, today,

\begin{abstract}
We revisit quantum state preparation of an oscillator by continuous linear position measurement. Quite general analytical expressions are derived for the conditioned state of the oscillator. Remarkably, we predict that quantum squeezing is possible outside of both the backaction dominated and quantum coherent oscillation regimes, relaxing experimental requirements even compared to ground-state cooling. This provides a new way to generate non-classical states of macroscopic mechanical oscillators, and opens the door to quantum sensing and tests of quantum macroscopicity at room temperature.
%\begin{description}
%\end{description}
\end{abstract}
%%TC:endignore
\maketitle
%\tableofcontents
Measurements are widely used to prepare quantum states, with applications ranging from quantum computing~\cite{riste_deterministic_2013}, to quantum sensing~\cite{cox_deterministic_2016,sayrin_real-time_2011} and the fundamentals of quantum mechanics~\cite{minev_catch_2019}. Continuous measurement of the position of a mechanical oscillator is particularly well studied, introducing a standard quantum limit to measurement precision~\cite{braginsky_quantum_1980} that has important consequences for gravitational wave detection~\cite{muller-ebhardt_quantum-state_2009,miao_achieving_2010} and other precision optomechanical sensors~\cite{brawley_nonlinear_2016,harris_minimum_2013}, and allowing feedback cooling towards the quantum ground state~\cite{wilson_measurement-based_2015,rossi_measurement-based_2018,doherty_quantum_2012}.
 
In the usual regime in which the rotating wave approximation (RWA) is taken for both the measurement and interactions with the environment, continuous measurement localizes the position and momentum of the mechanical oscillator equally~\cite{doherty_quantum_2012,bowen_quantum_2015}. This precludes the generation of non-classical states. Furthermore, even approaching the ground state requires a measurement speed that exceeds the thermalisation rate of the oscillator~\cite{clerk_back-action_2008,bowen_quantum_2015}. This has so far precluded the demonstration of ground state cooling at room temperature.

It has been understood for some time that sufficiently strong measurements can break the symmetry between position and momentum, allowing the preparation of quantum squeezed states of an isolated oscillator~\cite{doherty_feedback_1999}, and generating macroscopic entanglement in a gravitational wave interferometer in the free-mass limit~\cite{muller-ebhardt_entanglement_2008,muller-ebhardt_quantum-state_2009,miao_achieving_2010}. Here, we consider this problem in the general case for an oscillator that is in contact with a thermal bath. By applying an optimal estimation formalism that is valid outside the usual rotating wave approximation, we derive analytical expressions for the variances and covariance of the conditional state prepared by linear measurement. These expressions apply for arbitrary measurements and for mechanical systems ranging from the free-mass limit of gravitational wave detectors to high frequency mechanical oscillators.

We find that conditional quantum squeezed states can be prepared, and derive analytic criteria for the parameter regimes where this is possible. Our results show that quantum squeezing persists, even outside both the quantum backaction dominated regime and the regime of quantum coherent oscillation (QCO), where it has commonly been viewed that quantum dynamics are not possible~\cite{aspelmeyer_cavity_2014,tsaturyan_ultracoherent_2017,chan_laser_2011}, and that the squeezing is highly robust to both temperature and detection efficiency. Strikingly, the requirements are greatly relaxed, even compared to those usually required for ground state cooling. Using parameters already achieved in a range of optomechanical devices~\cite{rosenberg_static_2009}, we predict that quantum squeezing is possible for macroscopic low frequency oscillators at room temperature. This provides a pathway to practical quantum enhanced mechanical sensors and the possibility for tests of macroscopic quantum mechanics in a new regime  of mass and temperature~\cite{muller-ebhardt_entanglement_2008,muller-ebhardt_quantum-state_2009,miao_achieving_2010}.

\begin{figure}
	\includegraphics[width=\columnwidth]{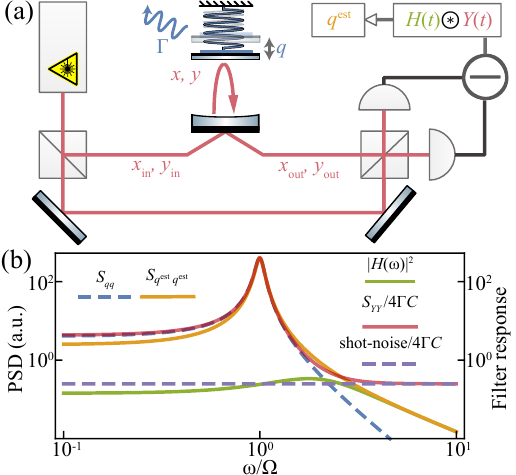}
	\caption{\label{fig:diagram}  Wiener filter for cavity optomechanical systems. (a) Schematic of proposed experiment. Mechanical motion (blue) is detected by homodyne measurement of the optical phase quadrature, $y_{\text{out}}$ (red). The Wiener filter ($H$) filters  the homodyne photocurrent ($Y$) providing the estimated position, $q^{\mathrm{est}}$ (orange). (b) Power spectral densities of the signals and the filter response function (green solid line) in the frequency domain. Position estimate: orange solid line; normalized optical readout: red solid line; normalized shot-noise: purple dashed line.}
\end{figure}

We consider the state of a mechanical resonator conditioned on continuous measurement of its position and weakly coupled to a thermal bath at temperature $T$, as shown in Fig.~\ref{fig:diagram}~(a). Together with oscillation at the mechanical resonance frequency $\Omega$, the dynamics of such a system involve a competition between the measurement conditioning, dissipation and noise introduced by the bath, and quantum backaction due to the measurement process~\cite{doherty_quantum_2012,bowen_quantum_2015,wilson_measurement-based_2015,rossi_measurement-based_2018,mason_continuous_2019,rossi_observing_2019}. Though our analysis is valid more generally, we consider measurement via an optical cavity, forming an optomechanical system. We describe the light by the dimensionless amplitude and phase quadratures, $x$ and $y$, respectively,  and the mechanical oscillator by the dimensionless position and momentum, $q$ and $p$. We normalize the variance of the zero-point fluctuations to unity, giving commutation relations $[x,y]=[q,p]=2i$. 
The linearized Hamiltonian, represented in a frame rotating on resonance with the cavity, is $\left. \hat H=\hbar  \Omega \left(q^2+p^2\right)/4+\hbar g xq\right.$~\cite{bowen_quantum_2015}. Here $g$ is the optomechanical coupling rate boosted by the coherent amplitude of the intracavity field. 

For a thermal bath with occupancy $\left. n_{\mathrm{th}} \approx k_{\text{B}} T/ \hbar \Omega \gg 1\right.$~\cite{kohen_phase_1997}, the observables in the Hamiltonian are governed by the Langevin equations~\cite{bowen_quantum_2015,giovannetti_phase-noise_2001}:
\begin{subequations}
\label{eqns:equ_motion}
\begin{align}
	\dot{x} & =-\frac{\kappa}{2}x+\sqrt{\kappa} x_{\mathrm{in}}
	\\
	\dot{y} & =-\frac{\kappa}{2}y+\sqrt{\kappa } y_{\mathrm{in}}-2g q
	\\
	\dot{q} & =+\Omega  p
	\\
	\dot{p}& =-\Omega  q-\Gamma  p+\sqrt{2\Gamma }p_{\mathrm{in}}-2g x,
\end{align}
\end{subequations}
where $\Gamma $ and $\kappa $ are the mechanical and optical energy decay rates, respectively; $x_{\mathrm{in}}$ and $y_{\mathrm{in}}$ refer to the optical vacuum noise inputs; and $p_{\mathrm{in}}$ is a thermal noise operator with power spectral density $S_{p_{\mathrm{in}}p_{\mathrm{in}}}=2n_{\mathrm{th}}+1$. The mechanical position $q$ is imprinted on the optical phase quadrature, allowing the mechanical state to be characterized via phase measurement, while the optical amplitude quadrature is imprinted on the mechanical momentum, introducing measurement backaction. We solve the equations of motion in the steady-state via Fourier transforming, with the convention $F(\omega) = \int_{-\infty}^\infty f(t)e^{i\omega t} \text{d} t$. We restrict the analysis to the unresolved-sideband regime ($\kappa \gg \Omega$), where the cavity field adiabatically follows the oscillator and the input, so that the optical cavity can be approximated to have flat frequency response.

The phase quadrature of the output optical field is given by  the input-output relation $y_{\mathrm{out}}=y_{\mathrm{in}}-\sqrt{\kappa}y$~\cite{gardiner_input_1985}. Detection of that quadrature yields a photocurrent $\left.Y=\sqrt{\eta}y_{\mathrm{out}}+\sqrt{1-\eta}y_{\mathrm{in}}'\right.$, with detection efficiency $\eta$ and an additional vacuum noise input $y_{\mathrm{in}}'$ included to model the effect of imperfect detection. This effects a linear, continuous measurement of the mechanical position. The optical quadrature $Y(t)$ satisfies the commutation relation $[Y(t),Y(t')]=0$~\cite{buonanno_signal_2002}. As a result the measured signal can be treated classically, allowing us to employ classical filtering to estimate the mechanical position by $q^{\mathrm{est}}=H(t)\circledast Y(t)$~ \cite{wiener_extrapolation_1964,jacobs_quantum_2014,doherty_feedback_1999}. The optimal choice of the filter $H(t)$ to minimize the uncertainty is the causal Wiener filter \cite{wiener_extrapolation_1964}, which we derive to be \footnote{Supplemental material} 
\begin{equation}
\label{eqns:filterq}
H(\omega)=A(1-iB\omega)\chi'(\omega)
\end{equation}
where  $\left. \chi'(\omega)=1/(\Omega '^2-\omega ^2-i \Gamma ' \omega)\right.$ is a modified mechanical susceptibility, and $\left. A=8\sqrt{\eta \Gamma ^3 C}n_{\text{tot}}\Omega ^2/( \Omega ^2+\Omega '^2) \right.$ and $\left.B=(\Gamma +\Gamma ')/(\Omega '^2-\Omega ^2+\Gamma^2  +\Gamma\Gamma')\right.$ are frequency-independent coefficients. The filter can be seen as a combination of $ \chi'(\omega)$ and its time-domain derivative, $-i \omega \chi'(\omega) $. The susceptibility is peaked at a resonance frequency $\left.\Omega' = \Br{16\eta \Gamma ^2 C n_{\text{tot}} \Omega ^2+\Omega ^4}^{1/4} \right.$, with decay rate $\left.\Gamma'  = \Br{-2\Omega ^2+\Gamma^2+2 \Omega '^2}^{1/2}\right.$, where we have introduced the optomechanical cooperativity $\left.C = 4g^2/\Gamma \kappa \right.$  and the effective total occupancy $\left.n_{\text{tot}} = n_{\mathrm{th}} + C + 1/2\right.$. As $C$ increases, both $\Omega'$ and $\Gamma'$ grow and the filter function becomes flat over a broad spectral range up to the resonance frequency $\Omega'$ (see Fig.~\ref{fig:diagram}~(b)).

Together with a similar filter to estimate momentum (derived in \cite{Note1}), the optimum for position filter in  Eq.~\ref{eqns:filterq}, produces a conditional state of the mechanical oscillator with uncertainty in both observables reduced beneath their intrinsic thermal-noise-limited values~\cite{doherty_quantum_2012, bowen_quantum_2015,brunelli_conditional_2019,rossi_observing_2019}. Since the measurement is linear, the conditional state of the oscillator is fully described by its conditional position and momentum variances and covariance~ \cite{wiseman_quantum_2009,jacobs_quantum_2014}. For our Wiener filters,  these are

\begin{subequations}
\label{eqns:covariancesolutions}
\begin{align}
  V_{\delta q \delta q}=&\frac{\Gamma' -\Gamma }{4 \eta C \Gamma }
  \\
  V_{\text{$\delta p$}\text{$\delta p$}}=&\frac{\text{$\Gamma' $} \left(\Gamma ^2-\Gamma  \text{$\Gamma' $}+\text{$\Omega' $}^2\right)-\Gamma  \Omega^2}{4 \eta C \Gamma    \Omega ^2}
  \\
  C_{\text{$\delta q$}\text{$\delta p$}}=&\frac{\Gamma ^2-\Gamma  \text{$\Gamma' $}+\text{$\Omega' $}^2-\Omega ^2}{4 \eta C \Gamma    \Omega },
\end{align}
\end{subequations}
where $\delta q = q-q^{\mathrm{est}}$ and $\delta p = p-p^{\mathrm{est}}$. From Eqns~\eqref{eqns:covariancesolutions}, it can be seen that four dimensionless variables describe the full behavior of the system: the thermal occupation $n_{\mathrm{th}}$, cooperativity $C$, measurement efficiency $\eta$, and quality factor $Q=\Omega/\Gamma$. The competition between measurement, oscillation and dissipation can be understood by comparing the measurement speed $\mu \equiv  C \Gamma$ to the oscillation frequency $\Omega$ and thermalisation  rate $\Gamma_{\text{th}} = n_{\text{th}} \Gamma$, leading to nontrivial conditions for the validity of the RWA and for quantum squeezing, as we will see later.

\begin{figure}
	\includegraphics[width=\columnwidth]{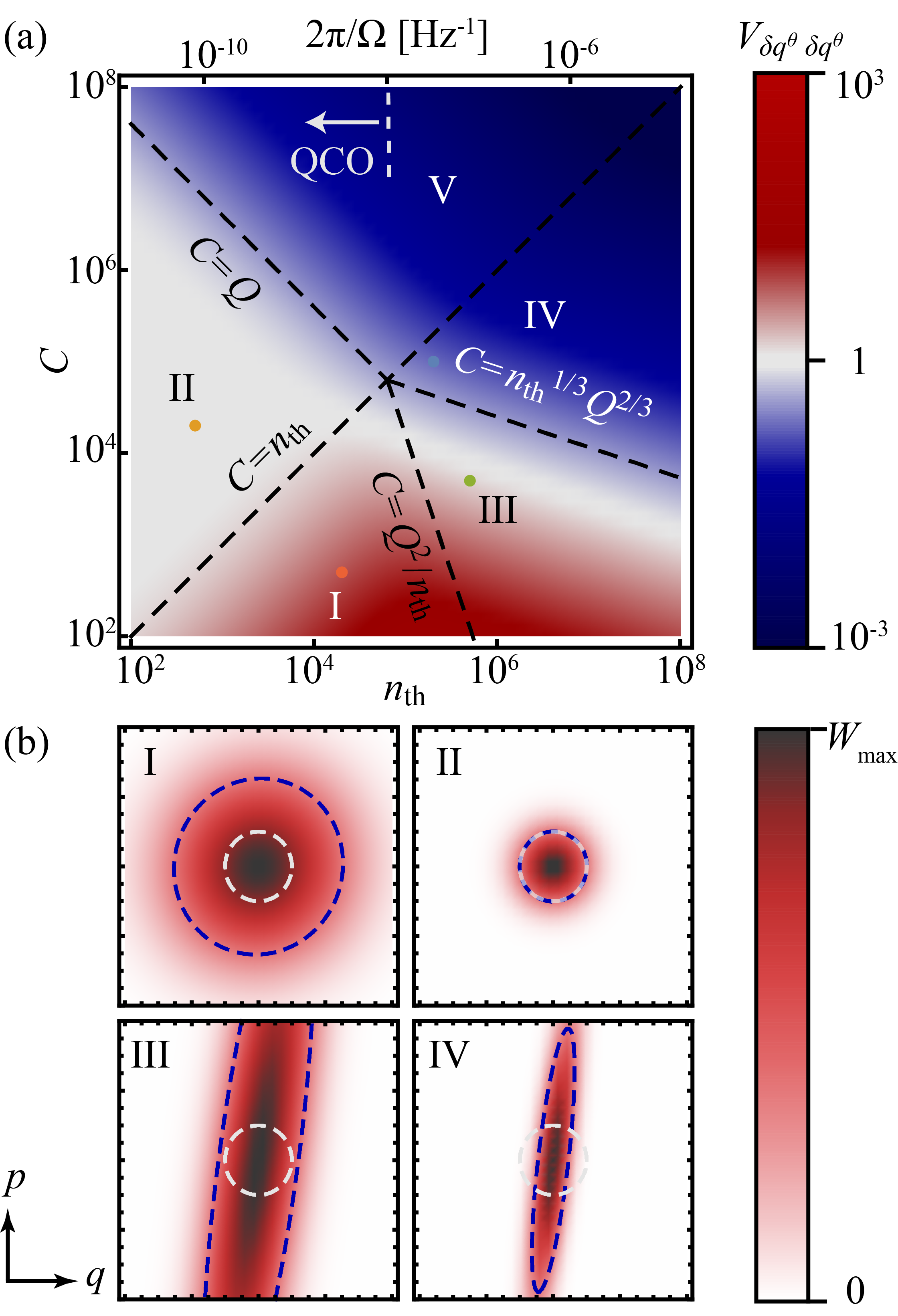}
	\caption{\label{fig:colormap} (a)~Characterization of the optimal variance $V_{\delta q^{\theta}\delta q^{\theta}}$ of the conditional state as a function of $n_{\mathrm{th}}$ (at fixed temperature; $n_{\mathrm{th}}\propto \Omega^{-1}$) and $C$ with $\eta=1$. Black dashed lines show thresholds for different measurement regimes. Colored dots I--IV show the parameters used to generate the Wigner functions $W(q,p)$ in Fig.~\ref{fig:colormap}~(b); white vertical dashed line separates the QCO regime (left) and non-QCO regime (right). (b) (I-IV) Wigner distribution of conditional states. Blue dashed ellipses: contour line of $W=e^{-1} \times W_{\text{max}}$; white dashed circles: ground state.}
\end{figure}

Outside  the regime of validity of the RWA, the covariance (Eq.~(\ref{eqns:covariancesolutions}c)) is non-zero, and the position and momentum are correlated. Consequently, there exists an optimal rotated quadrature with minimum variance. This optimal quadrature is given by 
$\left.\delta q^{\theta }=\delta q~ \text{cos$\theta $}+\delta p~ \text{sin$\theta $}\right.$ where the optimal angle $\theta=-\text{arctan}(\Omega /\Gamma')/2$. This angle approaches $-\pi/4$ as the measurement speed $\mu \rightarrow 0$ ($\Gamma' \ll \Omega$), and approaches $0$ as $\mu \rightarrow \infty$ ($ \Gamma' \gg \Omega$). In the latter case, the measurement tends towards a projective position measurement, where the conditional state approaches a position eigenstate. A relatively fast transition of the optimal angle occurs between these two limiting cases at around $\eta C n_{\mathrm{tot}}\sim Q^2$, which we find corresponds to the boundary of validity of the RWA. The breakdown condition $\eta C n_{\mathrm{tot}} > Q^2$ has a simple physical interpretation in the frequency domain, as the regime where the measured mechanical position power spectral density at $\omega=0$ exceeds the shot-noise. This is also the regime in which the bandwidth of the resolved mechanical motion exceeds the mechanical frequency. Deep within the non-RWA regime, we find that the purity of the state is $\mathcal{P}=\sqrt{{\eta  C} / {n_{\text{tot}}}}$ (see \cite{Note1}). For unity detection efficiency, the conditional state approaches a pure quantum state ($\mathcal{P}=1$) once the measurement speed exceeds the thermalisation rate, \textit{i.e.} $\mu \gg \Gamma_{\text{th}}$ (or equivalently $C \gg n_{\text{th}}$).

The variance of the optimal quadrature is shown as a function of $C$ and $n_{\mathrm{th}}$ in Fig.~\ref{fig:colormap}~(a) with unity detection efficiency. Here the bath is fixed at room temperature so that an increase in $n_{\mathrm{th}}$ corresponds to a decrease in the mechanical frequency. The conditional states fall into three broad categories, corresponding to variances larger than, equal to, or less than unity. These are depicted in red (thermal), white (ground), and blue (quantum squeezed), respectively. As we show below, these categories can be further subdivided into a total of five distinct measurement regimes (I--V) based on the interplay of $C$, $n_{\text{th}}$, and $Q$.

In regions I and II of Fig.~\ref{fig:colormap}~(a) the measurement is weak enough that multiple periods are required to resolve the mechanical motion, with equal amounts of information extracted about the position and momentum ($V_{\delta q \delta q}=V_{\delta p \delta p}$ and $C_ {\delta q \delta p}=C_ {\delta p \delta q} =0$)~\cite{doherty_quantum_2012,bowen_quantum_2015}. Consequently, the conditional state remains essentially symmetrical and is well-approximated by the RWA result, with ground state cooling achieved for $\mu \gg \Gamma_{\text{th}}$~\cite{rossi_measurement-based_2018,wilson_measurement-based_2015, mason_continuous_2019, rossi_observing_2019,doherty_quantum_2012,bowen_quantum_2015}. Reducing the variance further requires moving into a regime where  the RWA is invalid ($\left. \eta C n_{\mathrm{tot}} > Q^2\right.$). Three new regions emerge in this case (regions III--V in Fig.~\ref{fig:colormap}~(a)): region III exhibits classical squeezing, region IV exhibits impure quantum squeezing, and in region V the state of the oscillator approaches a pure quantum squeezed state in the limit of perfect detection efficiency. We see, therefore, that fast continuous measurements -- with measurement speed $\mu$ sufficiently large to invalidate the RWA -- are capable of producing conditional quantum squeezed states. We find the criterion for quantum squeezing is $C>n_{\text{tot}}^{{1}/{3}}Q^{{2}/{3}}/4\eta $. The free-mass limit where $\Omega \rightarrow 0$ corresponds to the very extreme of region IV. In that limit, our results can be shown to agree with predictions for the free-mass limit of gravitational wave detectors~\cite{muller-ebhardt_quantum_2009}. Wigner distributions of the conditional states for representative points in Fig.~\ref{fig:colormap}~(a) are computed using the covariance matrix (see \cite{Note1}), and shown in Fig.~\ref{fig:colormap}~(b).

By taking the appropriate limits of Eqns.~\ref{eqns:covariancesolutions}, expressions for the boundaries between the regions can be found. The boundary between region III and IV occurs at $\left.  C \sim n_{\text{tot}}^{{1}/{3}}Q^{{2}/{3}}/\eta\right.$, which asymptotes to $\left.  C \sim n_{\text{th}}^{{1}/{3}}Q^{{2}/{3}}/\eta\right.$ for $C\ll n_{\text{th}}$, as applies across most of the boundary. The boundary between regions IV and V occurs when the measurement speed equals the thermalisation rate ($\mu \sim \Gamma_{\text{th}}$, or equivalently $C \sim n_{\text{th}}$). The boundaries between regions I and III and between II and V correspond to the transition between validity and invalidity of the RWA, \textit{i.e.} $ C\sim Q^2/n_{\text{tot}}\eta$. Note that the right-hand side of this expression depends on $C$, through $n_{\text{tot}}$. This implicit dependence can be removed on the boundary between regions I and III by observing that $C \ll n_{\text{th}}$ for the majority of this boundary so that $n_{\text{tot}} \sim n_{\text{th}}$. The boundary then becomes $ C\sim Q^2/n_{\text{th}}\eta$. Similarly, the boundary between II and V can be simplified to $ C \sim Q /\eta$ by observing that $ C \gg n_{\text{th}}$ across most of this boundary.

Notably, both region IV and a portion of region V lie outside of the quantum coherent oscillation regime. That is, quantum squeezing is possible even when $Q<n_{\text{th}}$ (or equivalently $Qf<k_{\text{B}}T/2\pi\hbar$, with $f=\Omega/2\pi$). This inequality is satisfied everywhere to the right of the white dashed line in Fig.~\ref{fig:colormap}~(a). The presence of quantum squeezing in this regime violates a widely accepted `rule of thumb' in the optomechanics literature that quantum phenomena can only be observed when the thermal noise is sufficiently weak that the mechanical oscillator remains coherent for many mechanical cycles, \textit{i.e.} when $Qf$ is much \textit{larger} than $k_{\text{B}}T/2\pi\hbar$ \cite{aspelmeyer_cavity_2014,tsaturyan_ultracoherent_2017,chan_laser_2011}. Being able to generate conditional squeezing with $Q/n_{\text{th}} \ll 1$---outside of the QCO regime  \cite{khosla_quantum_2017}---significantly relaxes the experimental requirements for harnessing room temperature quantum effects.

Region IV of Fig.~\ref{fig:colormap}~(a) shows that quantum squeezing can even be achieved outside of the backaction dominated regime. In fact, the cooperativity required to achieve quantum squeezing decreases  at lower mechanical resonance frequencies even though this increases the thermal bath occupancy. As a result, continuous measurement and estimation provide an avenue towards the conditioning of non-classical states of low-frequency oscillators with requirements significantly relaxed compared to other techniques. Typically $\eta C>n_{\mathrm{th}}$ is required to prepare non-classical states \cite{clerk_back-action_2008,brunelli_conditional_2019,bowen_quantum_2015}. By contrast the protocol presented here only requires $\eta C>n_{\mathrm{th}}(Q/n_{\mathrm{th}})^{2/3}$. For example, for a room temperature mechanical oscillator with quality factor of $10^5$ and resonance frequency of 100~kHz, this relaxes the required cooperativity by a factor of $(n_{\text{th}}/Q)^{2/3} \sim 70$. Indeed, the required optomechanical measurement speed, $\mu=C \Gamma$, is highly insensitive to the mechanical decay rate and bath occupancy, scaling to the power of one third with both parameters. Since these parameters are typically much degraded at room temperature compared to cryogenic conditions, this is an attractive property for room temperature quantum optomechanics.

\begin{figure}
	\includegraphics[width=\columnwidth]{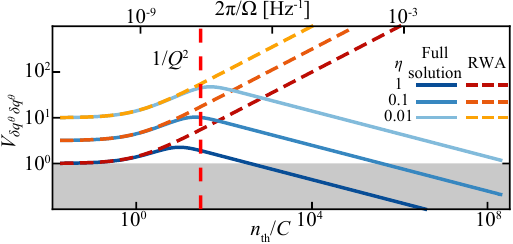}
	\caption{\label{fig:linecurve} Optimum variance of the conditional state for different measurement efficiencies ($\eta$) at fixed temperature $T=300~\text{K}$ and cooperativity $C=5\times10^3$ as a function of $n_{\text{th}}$  (or equivalently $\Omega^{-1}$), with and without the RWA.}
\end{figure}

The contrast between the non-RWA solution given here and the RWA is illustrated in  Fig.~\ref{fig:linecurve} for fixed $C$ and $T$ (note that for fixed $T$, $n_{\text{th}} \propto \Omega^{-1}$), and for various measurement efficiencies $\eta$. Beginning in the backaction dominated regime $n_{\mathrm{th}}/C \ll 1$, and in the regime of validity of the RWA ($ \eta n_{\mathrm{tot}} C \ll Q^2$), we see both solutions predict a constant variance $1/\sqrt{\eta}$ as expected. As the thermal occupancy increases and the system moves out of the backaction dominated regime the variances increase in both cases. Further increasing the thermal occupancy, the RWA solution predicts the conditional variance will increase monotonically. By contrast, the full solution predicts that the optimal squeezed variance will reach a peak at around $\eta C \sim  Q^2/ n_{\mathrm{th}}$, before reducing, eventually becoming squeezed below the level of the zero-point fluctuations, and approaching zero as $C \rightarrow \infty$. This occurs for all finite efficiencies, though the measurement speed required to achieve quantum squeezing increases as the efficiency becomes lower.

This effect hinges on the fact that both thermal and backaction noise couple directly to the oscillator's momentum, which takes a finite time to affect its position. Specifically, for short evolution times these two types of noise are suppressed, only influencing the mechanical position variance to third order (and above) in $\Omega t$~\cite{vanner_pulsed_2011, khosla_quantum_2017,bennett_quantum_2016}. The optomechanical measurement speed therefore need not be faster than the thermalisation rate, but only the rate at which noise couples into the position, relaxing the requirements for quantum squeezing at low mechanical resonance frequencies. Indeed, from this perspective the boundary between regions III and IV in Fig.~\ref{fig:colormap}~(a) may be qualitatively derived from the requirement that the measurement be able to resolve the zero-point motion of the oscillator in a time shorter than the time over which over which thermal and backaction noise of magnitude equal to the zero-point fluctuations enters the position of the oscillator. As a result of the increased noise suppression with decreasing mechanical frequency, there is a regime for which it is advantageous to move to a lower mechanical resonance frequency when attempting to condition a quantum squeezed state. Furthermore, it is seen in Fig.~\ref{fig:linecurve} that the protocol is able to produce a quantum squeezed state for any detection efficiency, albeit at the penalty of a higher requisite cooperativity. Lower detection efficiency results in a less pure state.

To explore the experimental feasibility of our scheme, we consider the experimental parameters in a spiderweb double disk device \cite{rosenberg_static_2009} which has $\Omega/2\pi=$ 694 kHz and single photon coupling strength  $g_0 /2\pi=$ 152 kHz. For $\eta=0.5$ a quantum squeezed state may be conditioned at room temperature with only $9\times10^3$ intracavity photons to satisfy \( C>n_{\text{th}}^{{1}/{3}}Q^{{2}/{3}}/4\eta\), corresponding to 675~nW of input optical power and a measurement speed of $\mu/2\pi=4.5 $ MHz. Intra-cavity phonon numbers and optical powers in this range are common in optomechanical systems (see e.g.~\cite{purdy_observation_2013}). As shown in \cite{Note1}, our protocol is robust to other technical noise, which can be included to derive new optimal filters. For example, if $1/f$ noise dominates shot-noise below $\Omega/5$ the cooperativity required to generate quantum squeezing is only increased by 43\%.

In summary,  we have theoretically explored continuous oscillator position measurement outside the rotating wave approximation and  including  interaction with a high temperature bath. We show that in the regime where the rotating wave approximation is invalid, quantum squeezing can be generated and that this can be achieved outside of the backaction dominated and quantum coherent oscillation regimes. This significantly relaxes the requirements to generate quantum states of macroscopic, low frequency oscillators at room temperature.


%\nocite{*}
%%TC:ignore
\begin{acknowledgments}
This research was supported primarily by the Australian Research Council Centre of Excellence for Engineered Quantum Systems (EQUS, CE170100009) and the Future Fellowship FT140100650. It was also supported by the Air Force Office of Scientific Research (FA9550-17-10397), the Commonwealth of Australia as represented by the Defence Science and Technology Group of the Department of Defence, and Lockheed Martin Corporation through the Australian Research Council Linkage Grant LP140100595. CM acknowledges support through an Australian Government Research Training Program Scholarship. We would like to thank A.C. Doherty, H.M. Wiseman, K.T. Laverick, and S. Khademi for useful discussion.
\end{acknowledgments}

%%TC:endignore

\section*{References}
%\nocite{*}
%\footnotesize \renewcommand{\refname}{\vspace*{-30pt}}
\bibliographystyle{apsrev4-2} % Tell bibtex which bibliography style to use

\clearpage

\pagebreak

\widetext
\begin{center}
	\textbf{\large Supplemental Material: Mechanical squeezing via fast continuous measurement}
\end{center}
%%%%%%%%%% Merge with supplemental materials %%%%%%%%%%
%%%%%%%%%% Prefix a "S" to all equations, figures, tables and reset the counter %%%%%%%%%%
\setcounter{equation}{0}
\setcounter{figure}{0}
\setcounter{table}{0}
\setcounter{page}{1}
\makeatletter
\renewcommand{\theequation}{S\arabic{equation}}
\renewcommand{\thefigure}{S\arabic{figure}}
\renewcommand{\bibnumfmt}[1]{[S#1]}
\renewcommand{\citenumfont}[1]{S#1}

\section{Wiener filter}

The causal Wiener filter is \cite{wiener_extrapolation_1964_S}
\begin{equation}
\label{eqns:filter1}
H(\omega)=\frac{1}{M_{Y}}\left[\frac{S_{qY}}{M_{Y}^*}\right]_{+},
\end{equation}
where the cross spectral density is defined by~\cite{clerk_introduction_2010_S,bowen_quantum_2015_S}
\[
S_{AB}(\omega)=\int^\infty_{-\infty}e^{i\omega t}\langle A(t)B(0)\rangle \text{d}t=\int^{\infty}_{-\infty} \frac{d\omega^\prime}{2\pi}\langle{A^{\dag}(-\omega)}{B(\omega\prime)}\rangle,
\]
and $M_{Y}$ is the causal spectral factor that satisfies $S_{YY}=M_{Y}M_{Y}^*$ and only has poles and zeros in the lower half of the complex plane.  $[...]_+$ denotes the causal part of the function. In general, any function may be separated into the sum of its causal and anti-causal parts \cite{wiener_extrapolation_1964_S}, which can be decomposed by factorising the poles of the denominator into the upper and the lower halves of the complex plane and finding the partial fraction decomposition. Expanding Eq.~\ref{eqns:filter1} using this procedure, we find the filter function:
\begin{equation}
H(\omega)=A(1-iB\omega)\chi'(\omega).
\end{equation}

The analogous filter function for momentum, 
\begin{equation}
H_{p}(\omega)=-\frac{A B}{\Omega} \left(\Omega^{2}+i \omega \frac{\Omega^{\prime 2}-\Omega^{2}}{\Gamma^{\prime}+\Gamma}\right) \chi^{\prime}(\omega),
\end{equation}

can be derived from Eq.~\ref{eqns:filter1} with the replacement $q\rightarrow p$.

The conditional power spectra of $\delta q$ and $\delta p$ can be determined by applying the Wiener filters to the measurement record. Those spectra provide the variances of the conditional state, given in Eqns.~(3) of the main text, using  \cite{bowen_quantum_2015_S}

\begin{equation}
V_{AB}=\int^{\infty}_{-\infty} \text{Re}\left\{S_{AB}(\omega)\right\}\frac{d\omega}{2\pi}.
\end{equation}

For example, the position variance of the conditional state is \cite{brown_introduction_1996_S}

\begin{equation}
V_{\delta q \delta q}=\int_{-\infty }^{+\infty}S_{\delta q \delta q}\frac{\text{d}\omega }{2\pi}.
\end{equation}
%
%Using the momentum filter function similarly furnishes the momentum variance $V\sb{\delta p\delta p}$.
Similarly, the covariance $C_{\delta q \delta p}$ is \cite{brown_introduction_1996_S}:
\begin{equation}
C_{\delta q \delta p}=\int_{-\infty}^{+\infty}\text{Re}\left\{S_{\delta q \delta p}\right\}\frac{\text{d}\omega}{2\pi}.
\end{equation}

In the case considered in this paper the conditional state is completely characterised by these (co)variances, because the conditional state is Gaussian. This is guaranteed by the fact that the system dynamics, loss, and measurement are all linear \cite{wiseman_quantum_2009_S, jacobs_quantum_2014_S}.

\section{Purity}
The purity of the conditional state is given by $\mathcal{P}=1/\sqrt{|\mathbb{V}|}$, where the covariance matrix is:
\begin{equation}
\label{eqns:covariance_matrix}
\mathbb{V}=\left(
\begin{array}{cc}
V_{\delta q \delta q} & C_ {\delta q \delta p} \\
C_ {\delta q \delta p} &  V_{\delta p \delta p} \\
\end{array}
\right).
\end{equation}
From Eqs~(3) of the main text, the determinant of the covariance matrix can be shown to satisfy $|\mathbb{V}| \geq 1$, ensuring that $P \leq 1$ as required for any physical quantum state.

\section{Conditional states in the presence of $1/f$ noise}
To verify that our protocol can be robust to technical noise, we here briefly consider the effects of classical laser phase noise on the conditional mechanical state. This is achieved by numerical calculations that can be extended to any excess noise spectrum.

\begin{figure}[h]
	\includegraphics[width=0.5\columnwidth]{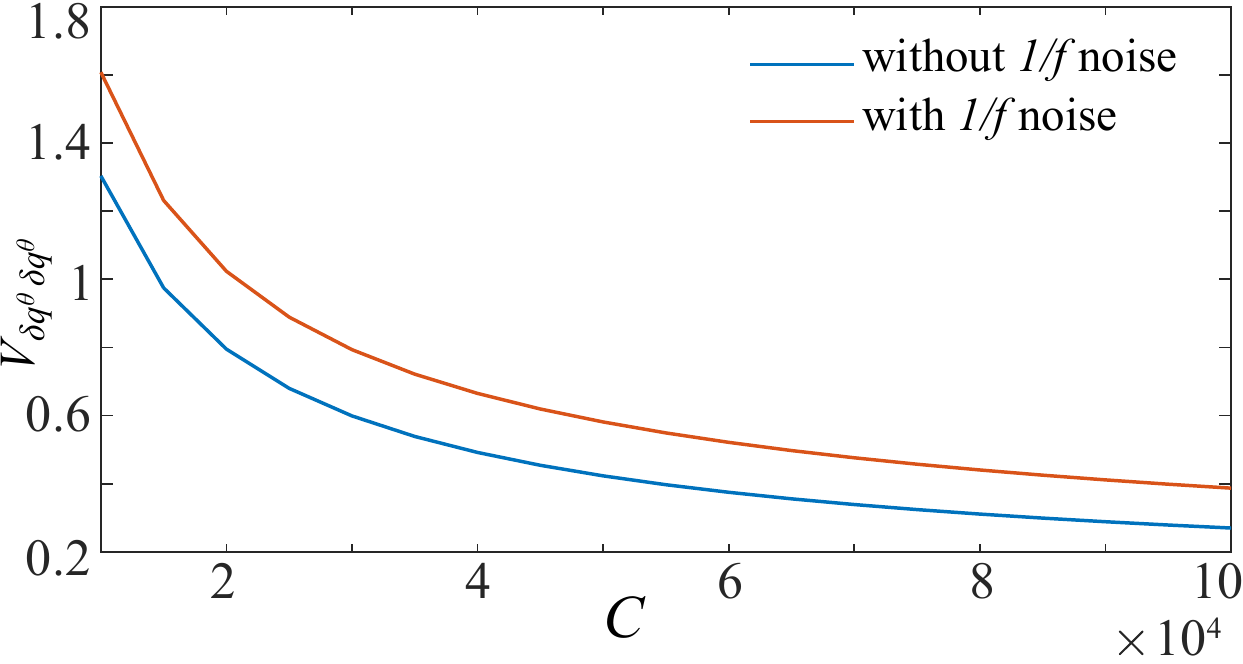}
	\caption{\label{fig:variance_pink_noise} The optimal variance as a function of cooperativity with and without excess $1/f$ noise. Quantum squeezing can be achieved in the presence of the $1/f$ noise, but requires 43\% more cooperativity .}
\end{figure}

\begin{figure}[h]
	\includegraphics[width=0.5\columnwidth]{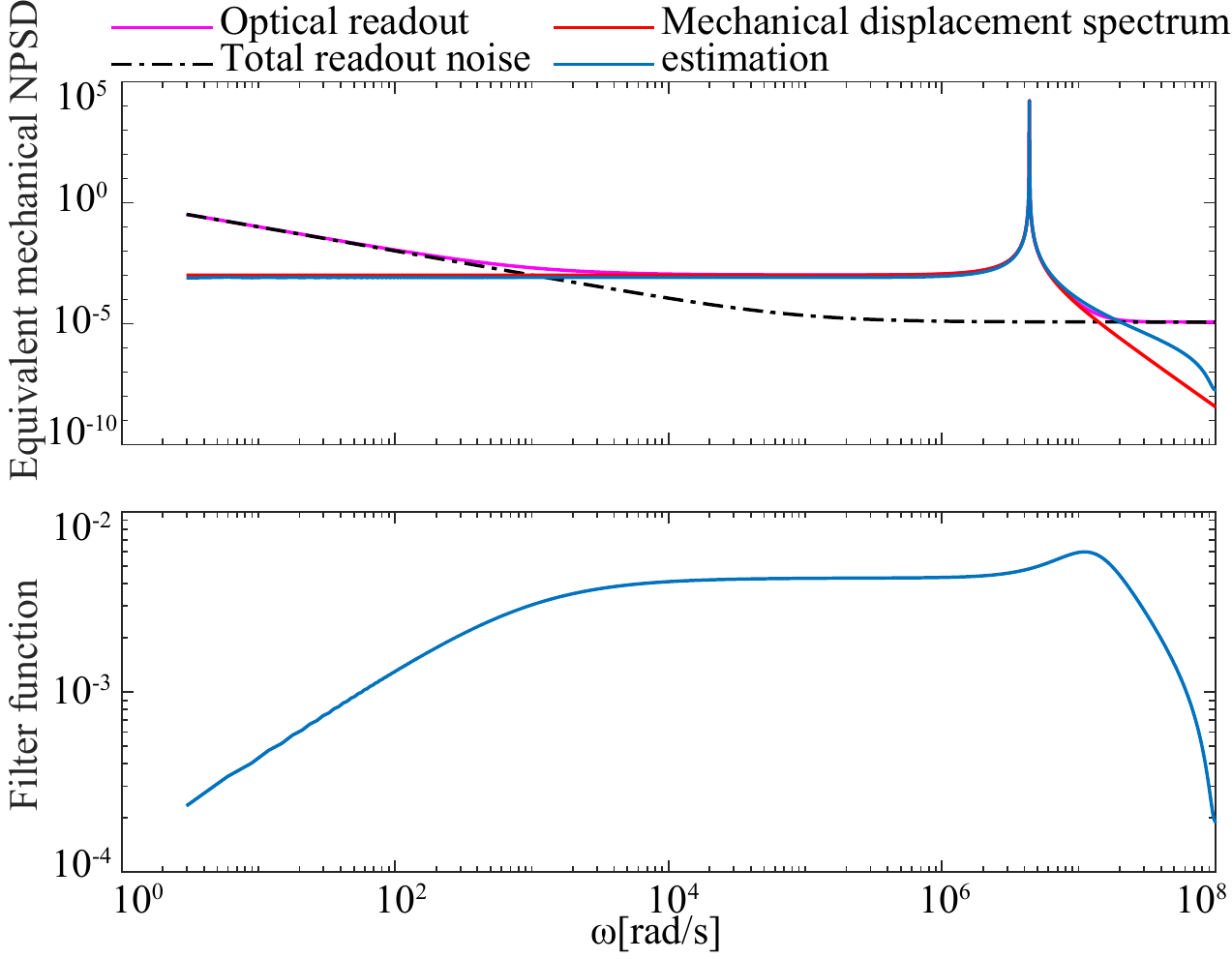}
	\caption{\label{fig:pink_spectrum}The equivalent mechanical noise power spectral density and position filter function spectrum. In the low frequency domain, the signal-to-noise ratio decreases in the presence of the excess noise. The filter compensates for this by weighting low frequencies less heavily than in the absence of $1/f$ noise.}
\end{figure}

Fig.~\ref{fig:variance_pink_noise} shows that quantum squeezing could still be achieved using existing devices \cite{rosenberg_static_2009_S}, even in the presence of $1/f$ excess phase noise (having a power spectral density $\propto \omega^{-1}$), as is commonly found in lasers. The optimal variance of the mechanical oscillator is shown as a function of measurement strength in the presence of pink noise with power spectrum $0.1\Omega/(\omega+0.1~\mathrm{rad/s})$ \footnote{We simulate the $1/f$ noise with an offset to avoid divergences at DC.} as shown in Fig.~\ref{fig:pink_spectrum}. This excess noise dominates from DC with 69~dB of excess noise until the corner frequency of $1.4\times10^5$~Hz by falling off at 3~dB per octave approximately.

\section*{References}
%\nocite{*}
%\footnotesize \renewcommand{\refname}{\vspace*{-30pt}}
\bibliographystyle{apsrev4-2} % Tell bibtex which bibliography style to use

\end{document}